# Perspectives of Women and Men Students and Faculty on Conceptual and Quantitative Problem-Solving in Physics from Introductory to Graduate Levels

Apekshya Ghimire* and Chandralekha Singh

Department of Physics & Astronomy, University of Pittsburgh, Pittsburgh, PA 15260, USA; clsingh@pitt.edu
* Correspondence: apg61@pitt.edu

**Abstract**

Developing expertise in physics requires appropriate integration and assimilation of physics and mathematics. Instructors and students often describe physics courses in terms of their emphasis on conceptual and quantitative problem-solving. For example, they may argue that a course emphasizes primarily conceptual over quantitative problem-solving or may emphasize equally on both depending on instructional context and assessment design. In this study, we investigated how students and instructors across different levels of physics instruction perceive the roles and development of conceptual and quantitative problem-solving in student learning and expertise development. Using departmental surveys administered at the beginning and end of each semester, we collected both Likert-scale and open-ended responses from students enrolled in introductory, upper-level undergraduate and graduate physics courses. These surveys assessed students' self-perceived skills, preferences, and perceptions of instructors and course emphasis. To complement student perspectives, we conducted interviews with instructors, using parallel questions adapted to reflect instructional goals and expectations. Our findings highlight patterns in how students and instructors prioritize conceptual and quantitative problem-solving across course levels, as well as alignment and misalignment between student and instructor perspectives. Also, although the questions were framed around conceptual versus quantitative problem-solving, we do not view them as mutually exclusive; rather we seek to understand perceived course emphasis and student expertise development from student and instructor points of view in a language commonly used in physics. These results can help shape teaching, course design, and assessment practices to better support the development of expert-like problem-solving skills in students in physics and related disciplines.

## 1. Introduction and Framework

*1.1. Problem Solving and Epistemological Framing*

Physics fundamentally uses mathematics to describe, explain, and predict the behavior of the natural world. The connection between physics and math is not incidental but central to how knowledge is developed and communicated [1-17]. Equations in physics are more than tools for calculation; they express relationships, define models of physical systems, and reveal patterns in nature [18-22]. Thus, learning physics involves more than just using mathematics as a tool to solve problems; it requires integrating physics and mathematics seamlessly to predict and explain physical phenomena [23-28]. As students progress, they must learn to use mathematics not only as a tool for calculations but to structurally integrate, evaluate and make sense of physical ideas [29-39]. Understanding this intertwined relationship between physics and mathematics is central not only for students, who are striving to develop physics expertise, but also for physics instructors so that they can support student learning at all levels, in introductory- to graduate-level courses.

Research shows that the difference between experts and novices in physics problem-solving involves how knowledge is organized and how it is activated while solving problems. Experts organize knowledge around deep conceptual structures, enabling them to flexibly integrate physics principles with appropriate mathematical representations [20, 40]. Novices, by contrast, often rely on surface features, which can lead them to engage in "plug and chug" approaches. Chi et al.'s work on expert–novice differences in problem solving [40] suggests that experts tend to categorize problems based on underlying physics principles, while novices focus on surface features, e.g., grouping problems as "inclined plane problems", "pulley problems" or "spring problems". In physics, this distinction often manifests in how students and faculty perceive problem solving: students may approach tasks as exercises in searching for and applying known equations, whereas instructors emphasize reasoning from fundamental principles and the relationships among physical quantities. These differing orientations influence how each group interprets what it means to "solve" a problem and what skills are most valued in physics learning.

Complementing ontological framing, some researchers have focused on epistemological framing [26, 41, 42], which highlights how student beliefs about the epistemology of physics problem-solving and the implicit expectations students hold about it affects whether they focus on a systematic approach that includes conceptual analysis and planning of a problem or launch into searching for a formula and plugging numbers. When instructors present worked examples that focus primarily on mathematical steps, it reinforces students' view that success in physics problem-solving means being able to find the most likely formulas and do plug and chug rather than reasoning about physical mechanisms systematically. This framing is reinforced through assessments that value plug and chug approaches and correct answers over explication of effective problem-solving strategies and articulated reasoning. Students quickly learn that if the final solution earns full credit, steps such as analyzing, planning, or reflecting upon and evaluating solutions are not that important.

At the heart of these synergistic framings is the role of mathematics within physics problem-solving. Researchers in both physics education and philosophy of science have long emphasized the inseparable connection between math and physics. Tzanakis [1], for example, argues that mathematics should not be treated simply as a computational tool but as intrinsic to the structure of physics itself that affords and enhances interpretation. Thus, he describes the relationship between physics and mathematics as a two-way inseparable process. Similarly, Redish notes that mathematics in physics often differs from that in math classes [43-45] as symbols and equations get their meaning from the physical ideas they represent and not just from the underlying mathematical rules. Pietrocola [46] likewise describes mathematics as the "structural language" of physics, e.g., as a way of organizing and expressing physical thought.

Building on these ideas, Uhden and colleagues developed a framework that helps to understand how mathematics is used in physics [18, 47]. According to their framework, math plays two roles in learning physics. The first is the technical role, such that students use mathematics to perform calculations, solve equations, and follow procedures. The second is the structural role, which involves math being used to build models that describe relationships between physical quantities and support reasoning. In this view, math is not just a skill set students use to get the right answer for physics problems; it is a way of thinking and making sense of physical ideas. Our study is inspired by these synergistic frameworks. We are especially interested in how students and instructors understand and use the structural role of mathematics in physics learning and problem-solving using the language of conceptual and quantitative problem solving prevalent in physics.

In physics education, conceptual and quantitative problem-solving are often discussed as distinct approaches. Conceptual problem-solving involves understanding the underlying ideas in a physical situation and identifying which principles apply, e.g., on conceptual physics inventories [48-55], while quantitative problem-solving focuses on performing symbolic or numerical calculations [56, 57]. However, conceptual and quantitative problem-solving are intricately connected and physicists integrate the two to make sense of physical phenomena [37, 58-61].

With these issues in mind, we use the terms "conceptual" and "quantitative" problem-solving in our study, although we recognize that they are fundamentally inseparable. We see them as two sides of the same coin, i.e., both need to be integrated structurally for meaningful learning and expert problem-solving in physics. For example, applying Newton's Second Law requires understanding the situation conceptually, drawing diagrams, identifying individual forces, and translating it into mathematical form. Even though integrating both types of reasoning are important for physics problem-solving, students often struggle to coordinate them [62-68]. In some cases, students rely mostly on equations and procedures, solving problems correctly without fully understanding the physics behind them [69-74]. In other cases, students may be able to describe concepts but have difficulty coordinating the math needed to apply those ideas in physics problems [38, 63, 65, 73, 75]. These challenges can happen at all levels, from introductory courses to graduate-level classes while solving problems.

*1.2. Cognitive Apprenticeship Model*

The Cognitive Apprenticeship Model [76] provides a useful framework for understanding how instructional practices can either reinforce or challenge these perceptions and plug and chug approaches to problem-solving. This model broadly includes three components: modeling, coaching and scaffolding, and weaning. In the modeling phase, instructors demonstrate expert thinking and problem-solving processes, making their expert-like approach visible to learners. Coaching and scaffolding involve providing guidance, prompt feedback and support while students work on problems to develop effective problem-solving skills. Finally, weaning refers to the gradual removal of support, encouraging learners to develop self-reliance in problem-solving. It is essential that conceptual aspects of problem-solving remain explicitly valued alongside quantitative aspects in an integrated manner so that this integration is seen as the defining feature of expert problem-solving.

In traditionally taught typical physics courses, physics instructors often emphasize only the quantitative aspects in detail. During modeling phase, instructors typically do not explicitly demonstrate (e.g., by writing down) how to do a conceptual analysis and planning of the solution and what students see projected on the board are quantitative steps in the implementation of solutions. Although conceptual aspects of problem-solving are embedded within these examples, students often overlook them since these aspects are not made explicit. Coaching and scaffolding are usually minimal in a typical course, and after lectures, students are weaned and asked to solve homework problems independently at home. Thus, without guidance, many students may continue to view physics problem-solving as primarily quantitative.

Furthermore, in most traditional physics courses, homework, quizzes, and exams focus primarily on symbolic or numerical problems with only the implementation step, i.e., showing quantitative steps being important for grades. Thus, other phases of problem-solving such as conceptual analysis, planning and evaluation, and reflection on the problem-solving process are overlooked. In particular, a holistic physics problem-solving approach, e.g., requiring students to use ASPIRE heuristic (i.e., Analyze and Sift through the problem, Plan the solution, Implement the plan, Reflect on and Evaluate the solution process to learn from the problem-solving process), is not emphasized and students are not required to do systemic problem-solving for receiving full grade. Such an emphasis on a structured approach to problem-solving would align with Uhden et al.'s (2012) framework and others discussed in the preceding section, which emphasize that mathematics and physics are deeply interconnected; mathematical formulations gain meaning through conceptual aspects of problem-solving. Explicitly rewarding all stages of the ASPIRE heuristic will signal to students that expert-like problem-solving involves more than finding and manipulating equations.

*1.3. Gender Difference in Physics Education*

The fact that traditional physics courses primarily emphasize quantitative aspects of problem-solving in modeling and assessment has important implications for gender equity in physics learning. Prior research suggests that gender disparities tend to be more pronounced on conceptual inventories than on final exams primarily requiring symbolic or numerical answers [77, 78]. This difference may stem not from ability but from gender differences in physics self-efficacy and identity as well as stereotype threat. Physics is a discipline that is dominated by men. For example, stereotypes about who belongs in physics and can succeed in it and the association of brilliance, e.g., in physics, with men [79-81] can negatively influence how women students engage in problem-solving, interpret classroom expectations, and assess their own capabilities [82-87]. Women in physics often have stereotype threat (fear of confirming a stereotype about their group), which can undermine their confidence and performance even when their abilities are comparable to their men peers [86, 88]. Such stereotype threat [85, 89-91] can impair performance by increasing anxiety and reducing working memory capacity during problem-solving [92, 93]. The lack of visible role models may further reinforce these barriers [94-96], making it harder for women to see themselves as capable of excelling in physics.

Self-efficacy, which is defined as a person's belief in their ability to successfully complete a specific task [97-99], plays a central role in this process. In physics, self-efficacy reflects a student's confidence in solving physics problems. Physics identity refers to the extent to which students see themselves as "a physics person", e.g., whether they can excel in physics courses [100-102]. Bandura's [97] four sources of self-efficacy (mastery experiences, vicarious experiences, verbal persuasion and psychological state) help explain why women may have lower self-efficacy in physics, especially in conceptual tasks that are not emphasized explicitly throughout by instructors. When students have few role models (related to vicarious experience) succeeding in physics [103, 104], their self-efficacy in physics is weakened. Also, stereotype threat heightens anxiety especially during conceptual problem-solving, which is typically not emphasized explicitly by physics instructors, undermining the emotional and physiological sources of self-efficacy [105, 106]. Thus, fewer relatable examples of success and increased anxiety make it challenging for women to feel confident while solving physics problems. When classroom and assessment structures emphasize quantitative over conceptual reasoning, women's experiences may result in lower self-efficacy and a weaker identity particularly pertaining to the conceptual aspects of physics problem-solving. This can shape how students engage with physics tasks, and how they interpret their performance and skills.

Therefore, equitable physics learning environments must intentionally strengthen these intersecting dimensions of cognition and psychological constructs. Carefully modeling and rewarding via grade incentives conceptual aspects of problem solving, e.g., through ASPIRE heuristic, can make the implicit processes of expert problem-solving visible to all students and emphasize its importance. When conceptual aspects of problem-solving are explicitly valued, assessed and developed throughout a physics course, it can support self-efficacy and identity development, particularly for those historically underrepresented in the discipline.

Building on these foundations, the present study investigates how students and instructors view the connection between physics and mathematics, particularly through the lens of conceptual and quantitative problem-solving since these terms are prevalent in the language commonly used in physics education. We examine gender differences in how students rate their conceptual and quantitative problem-solving skills as well as how they perceive what they want from the course, what their instructors emphasize, and what actually happens during instruction. By examining how students define these terms, how they evaluate their own abilities along these dimensions, and how they interpret the goals and practices of their courses, we hope to better support expert-like reasoning and problem-solving in physics. Our work spanning introductory to graduate level physics courses aims to highlight the importance of helping all students regardless of gender, integrate conceptual and quantitative problem-solving and reasoning to develop expertise in physics, and internalize that they are deeply connected ways of making sense of the physical world.

## 2. Methodology

*2.1. Research Design*

In this study, primarily quantitative survey data (with some open-ended written responses) from students across introductory, upper-level undergraduate and graduate physics courses and qualitative interview data from faculty were analyzed separately and then interpreted together to provide a comprehensive understanding of conceptual and quantitative problem-solving in physics. This approach allowed for direct comparison between quantitative patterns (such as gender differences in self-assessment) and qualitative insights from faculty regarding instructional emphasis and classroom practices. The design integrated cross-sectional student survey data from multiple course levels with semi-structured interviews from faculty to provide complementary perspectives. Quantitative analyses focused on descriptive statistics and trends across course levels, while qualitative analyses explored the reasoning behind those patterns.

*2.2. Course Context and Participants*

Data were collected over four consecutive academic years, encompassing four fall semesters and four spring semesters. The fall semesters were the on-semester offerings for algebra and calculus-based physics 1, while the spring semesters were the on-semester offerings for physics 2 courses. These semesters were chosen to ensure consistent student enrollment in these courses. Some students may appear in both datasets if enrolled in both course types. For upper-level undergraduate and graduate-level courses, we combined data from both the fall and spring semesters. The upper-level undergraduate data includes multiple courses spanning the second to fourth year of the program, which may result in some students appearing more than once in the dataset. In addition, seven faculty members were individually interviewed to capture instructor perspectives on conceptual and quantitative problem-solving, instructional emphasis, and assessment practices.

Participant selection was based on course enrollment. All students enrolled in the identified courses were invited to complete the surveys, and participation was voluntary. Students received extra credit as an incentive for completing the survey. Faculty participants were selected to represent a range of teaching experience, course levels, and instructional styles. Table 1 summarizes all courses included in the study, categorized by level and the number of women and men who responded to the survey questions: (CQ1) what students wanted the course to emphasize (conceptual, quantitative or both), (CQ2) what they thought their instructor wanted, and (CQ3) what they felt the course actually emphasized in the pre- and post-survey.

**Table 1.** Summary of courses included in the study, their level, primary focus and number of women and men who responded to the following survey questions in each course: (CQ1) what students wanted the course to emphasize (conceptual, quantitative, or both), (CQ2) what they thought their instructor wanted, and (CQ3) what they felt the course actually emphasized, in the pre- and post-survey. [Electricity and Magnetism is abbreviated as E&M in the table below].

| Courses | Level | Primary Focus | Number of Participants | | | |
| --- | --- | --- | --- | --- | --- | --- |
| | | | Pre-Survey | | Post Survey | |
| | | | Women | Men | Women | Men |
| Algebra-based Physics 1 | Introductory | Mechanics, Kinematics and Dynamics | 1552 | 824 | 1795 | 872 |
| Algebra-based Physics 2 | Introductory | E&M, Optics | 1545 | 867 | 1427 | 768 |
| Calculus-based Physics 1 | Introductory | Mechanics, Kinematics and Dynamics | 1079 | 1647 | 1048 | 1387 |
| Calculus-based Physics 2 | Introductory | E&M, Optics | 838 | 1327 | 668 | 1110 |
| Upper-Level Undergraduate Physics | Undergraduate | Advanced Mechanics, E&M, Quantum | 92 | 315 | 69 | 247 |
| Graduate-Level Physics | Graduate | Dynamical Systems, E&M, Quantum, Thermodynamics | 22 | 43 | 18 | 36 |

Tables 2 and 3 present the number of women and men students in algebra and calculus-based courses who self-assessed their conceptual and quantitative problem-solving skills in the pre- and post-surveys, respectively.

**Table 2.** Number of women and men who rated their conceptual and quantitative problem-solving skills in introductory level courses in the pre-survey.

| | Conceptual Skills | | Quantitative Skills | |
|---|---|---|---|---|
| | **Women** | **Men** | **Women** | **Men** |
| Algebra based Physics 1 | 512 | 285 | 516 | 280 |
| Algebra based Physics 2 | 404 | 187 | 407 | 189 |
| Calculus based Physics 1 | 262 | 485 | 260 | 482 |
| Calculus based Physics 2 | 156 | 324 | 157 | 324 |

**Table 3.** Number of women and men who rated their conceptual and quantitative problem-solving skills in introductory level courses in the post-survey.

| | Conceptual Skills | | Quantitative Skills | |
|---|---|---|---|---|
| | **Women** | **Men** | **Women** | **Men** |
| Algebra based Physics 1 | 514 | 239 | 520 | 243 |
| Algebra based Physics 2 | 310 | 115 | 302 | 120 |
| Calculus based Physics 1 | 177 | 312 | 176 | 312 |
| Calculus based Physics 2 | 115 | 227 | 113 | 227 |

Table 4 shows the number of students in undergraduate and graduate-level courses who self-assessed their conceptual and quantitative problem-solving skills in the pre- and post-surveys.

**Table 4.** Number of students who rated their conceptual and quantitative problem-solving skills in physics undergraduate and graduate-level courses in the pre- and post-survey.

| | Conceptual Skills | | Quantitative Skills | |
|---|---|---|---|---|
| | **Pre** | **Post** | **Pre** | **Post** |
| Undergraduate Upper Level | 89 | 130 | 91 | 130 |
| Graduate Level | 34 | 23 | 36 | 24 |

*2.3. Survey Instruments*

We investigated college students' perspectives on conceptual and quantitative problem-solving through a departmental survey administered at both the beginning and end of each semester. As shown in Table 5, the survey included 9 questions related to conceptual and quantitative problem-solving. Some questions, particularly those asking students to define these terms and describe strategies to future students in physics courses to develop them, were open-ended. Others involved self-assessments of conceptual and quantitative problem-solving skills using a 7-point Likert scale ranging from Very Poor to Exceptional. Self-assessment questions related to student ratings of conceptual and quantitative problem-solving skills administered in two consecutive fall semesters and one spring semester are analyzed only for those specific semesters.

Table 5. All the survey items (open-ended and Likert scale) related to conceptual and quantitative problem-solving given to students at introductory, upper level and graduate level physics courses.

| Categories | Survey Items |
|---|---|
| Definitions and strategies to develop skills (open-ended) | What does conceptual problem-solving in physics mean to you? How would you know that you have good conceptual problem-solving skills in physics? |
| | Write down three tips for a peer who is struggling to improve their conceptual problem-solving skills in this physics course. |
| | What does quantitative problem-solving in physics mean to you? How would you know that you have good quantitative problem-solving skills in physics? |
| | Write down three tips for a peer who is struggling to improve their quantitative problem-solving skills in this physics course. |
| Self-assessment of the skills (7-point Likert scale) | How would you rate your conceptual problem-solving skills in this physics course? |
| | How would you rate your quantitative problem-solving skills in this physics course? |
| Students' preferences (5-point Likert scale) | How do/did you want this course to improve your conceptual problem-solving skills in physics vs. quantitative problem-solving skills? (do—pre survey, did—post survey) |
| Instructors' emphasis (5-point Likert scale) | How do you think your instructor wants/wanted this course to improve your conceptual problem-solving skills in physics vs. quantitative problem-solving skills? (wants—pre survey, wanted—post survey) |
| Actual course emphasis (5-point Likert scale) | How do you think this course actually will improve/improved your conceptual problem-solving skills in physics vs. quantitative problem-solving skills? (will improve—pre survey, improved—post survey) |

The survey questions were worded in terms of "conceptual vs. quantitative" problem-solving, to use the language prevalent in physics and understand physics students' and instructors' views on these and whether they tend to focus more on one over the other. To investigate this, we asked students about their preferences for what they wanted to learn and get out of a physics course, their perceptions of the instructor's emphasis, and the actual emphasis in their courses using a 5-point Likert scale.

We also conducted individual interviews with seven faculty members from the physics department to investigate their perspectives on conceptual and quantitative problem-solving, as well as their perceptions of students' views on this topic. The interview questions were closely aligned with those in the student survey (shown in Table 5) but were adapted to reflect the instructor's role, asking what they emphasize in teaching and how they perceive student preferences across course levels. Faculty also discussed assessment emphasis and the proportion of students they expected to value conceptual over quantitative focus. The complete set of faculty interview questions used in this study is provided in Appendix A for reference.

*2.4. Data Analysis*

For clarity in data presentation, quantitative data were analyzed using descriptive statistics, with Likert-scale responses collapsed into broader categories. We grouped the responses for self-assessment items into four categories: Poor (combining Very Poor and Poor), Fair, Good (combining Good and Very Good), and Excellent (combining Excellent and Exceptional). We then calculated and compared the percentages of women and men who rated their conceptual and quantitative skills in each category, presenting the results in bar graphs for clarity.

Similarly, we grouped the responses related to students' perspective on whether the courses were conceptual or quantitative into three broader categories: Conceptual (combining Mostly conceptual problem-solving skills, and More conceptual than quantitative problem-solving skills but improvement on both), Both Equally, and Quantitative

(combining More quantitative than conceptual problem-solving skills but improvement on both, and Mostly quantitative problem-solving skills). In addition, each of these questions were followed by an open-ended prompt asking students to explain their reasoning, allowing us to gather qualitative insights into their interpretations and justifications.

Both researchers independently read and then collaboratively reviewed open-ended responses, discussing patterns and themes that emerged. Together, they agreed that students provided thoughtful and insightful responses, and selected quotes that were representative of the broader population or echoed ideas expressed by many students. The instructor interviews were transcribed (one interview was not recorded due to glitches but interviewer took extensive notes immediately after) and analyzed in the same manner with A.G. taking the lead to code and categorize quotes and iterating based upon discussions with C.S., allowing for direct comparison between student and faculty perspectives. Also, only the first author did all quantitative data analysis including the tallying of instructor responses for Table 8.

To investigate gender differences statistically, we calculated effect sizes with 95% confidence interval (CI) and performed $t$-tests on students' self-ratings of conceptual and quantitative skills. For introductory algebra and calculus-based courses, $t$-tests were conducted separately on pre-survey and post-survey ratings to determine whether statistically significant differences existed between women and men. For upper-level undergraduate and graduate courses, the focus was on pre-to-post changes within each student group, and $t$-tests were used to examine whether students' conceptual and quantitative ratings significantly changed over time. These analyses allowed us to assess how instructional experiences influenced students' perceptions across different course levels.

*2.5. Instrument Validation*

To ensure that the survey and interview instruments were interpreted as intended, we conducted multiple rounds of validation and refinement. During a summer semester, we tested two alternative versions of the student survey with students enrolled in introductory physics courses to examine whether question format influenced interpretation. The first version used the question-based format shown in Table 5, for example: "How did you want this course to improve your conceptual problem-solving skills in physics vs. quantitative problem-solving skills?" The second version used statement-based items rated on a 7-point Likert scale (from Strongly Disagree to Strongly Agree), such as "I wanted more conceptual focus than quantitative focus in this course" or "I wanted a balance between conceptual and quantitative focus in this course." The survey responses showed no significant differences in student responses across the two versions, indicating that both formats captured the same underlying perceptions, and students interpreted the two versions in a similar manner. Based on these findings, we retained the question-based format for consistency and clarity in subsequent semesters. To further validate the instrument, we also conducted semi-structured interviews with 13 physics students across all levels. The interviews explored how students understood the constructs of conceptual and quantitative problem-solving and whether their interpretations aligned with the survey items. Students' definitions and reasoning patterns closely mirrored those from the larger dataset, supporting the content validity of the survey questions. Moreover, five graduate students specializing in physics education research reviewed the survey to ensure clarity and confirm that the questions conveyed the intended meaning.

In addition to student validation, faculty members were also involved in refining both the survey and interview instruments. During the interview phase, faculty reviewed the survey items for clarity, relevance, and alignment with instructional practice. Their feedback guided iterative revisions to improve item phrasing and ensure that key aspects of conceptual and quantitative reasoning were appropriately represented. After each faculty interview, the researchers evaluated the feedback and adjusted the interview questions such as rewording items or adding probes to capture instructors' perspectives on assessment and course emphasis. This iterative approach continued until responses became consistent across participants, resulting in a well-calibrated set of survey and interview instruments that were both contextually appropriate and methodologically robust.

*2.6. Ethical Standards*

The study was conducted in accordance with the Declaration of Helsinki and approved by the University's Institutional Review Board. The research was approved as exempt by the university's Institutional Review Board and written informed consent requirement was waived. Verbal consent was obtained for individual interviews with each participant.

## 3. Research Questions

The purpose of this study was to examine how students across different levels of physics education, introductory, upper-level undergraduate, and graduate level perceive and define conceptual and quantitative problem-solving, and how these perceptions relate to instructional practices and course contexts. The study aimed to identify patterns in how students value, define, and self-assess these two types of problem-solving skills, and to explore whether faculty emphasize one over the other in their teaching.

Guided by this purpose, the study addressed the following research questions:

**RQ1.** *How do introductory students define conceptual and quantitative problem-solving in physics, how do they recognize when someone is good at these, and what strategies do they recommend to peers for improving conceptual and quantitative problem-solving skills?*

**RQ2.** *How do introductory women and men students' self-assessed conceptual and quantitative problem-solving skills differ by gender across various levels of physics courses?*

**RQ3.** *To what extent do upper-level undergraduates and graduate students' self-assessments of conceptual and quantitative problem-solving skills change from the beginning to the end of the course, and how does this vary across course levels?*

**RQ4.** *How do women and men students' preferences for instructional emphasis (conceptual vs. quantitative) compare with their perceptions of what their instructors emphasize and what the course actually emphasizes across different levels of physics courses at the beginning and end of the semester?*

**RQ5.** *How do faculty members perceive the alignment between students' instructional preferences, their own teaching goals, actual course implementation, and assessment focus in physics courses pertaining to conceptual and quantitative problem-solving?*

## 4. Results

**RQ1. How do introductory students define conceptual and quantitative problem-solving in physics, how do they recognize when someone is good at these, and what strategies do they recommend to peers for improving conceptual and quantitative problem-solving skills?**

Introductory physics student responses were remarkably consistent across all courses and levels, indicating a shared understanding of conceptual and quantitative problem-solving in physics. Students described conceptual problem-solving skills in physics as the ability to understand and apply the underlying principles of physical phenomena without relying solely on numbers or equations. For many, it means being able to visualize what's happening in a scenario, explain the reasoning behind formulas, and connect different concepts across contexts. Some noted that being able to teach or explain a concept to others, especially using visuals or real-world examples, was a strong indicator of conceptual mastery. One student wrote, *"Conceptual problem-solving means having the capacity to describe to another person how to correctly solve a hypothetical problem and provide a real-life application. If I were able to explain how to find the volume of a sphere, without the use of values, for example, I would say I have a good conceptual understanding of that problem."* Rather than simply memorizing equations, many students at all levels emphasized the importance of understanding why those equations work and how to apply them thoughtfully in unfamiliar situations as conceptual problem-solving. For example, another student noted, *"I believe that it is the ability to not just remember equations but being able to apply them to new concepts or completely different problems. I would know that I have good conceptual problem-solving skills in physics if I were able to sort of "connect-the-dots" between different concepts and be able to work through a problem that I have not seen before."*

To help peers improve their conceptual skills, students recommended a variety of strategies centered around active learning and consistent engagement with the course materials. Common tips included reading the textbook, attending office hours, watching explanatory videos, and drawing diagrams to visualize problems. Many emphasized practicing by explaining concepts aloud, redoing problems, and applying physics to real-life contexts. Many students stated that truly understanding what a question is asking and being able to break it down logically was key to success in conceptual problem-solving. Overall, they viewed conceptual problem-solving as essential not only for tackling physics problems but for building a deeper, lasting understanding of the subject.

Students defined quantitative problem-solving skills in physics as the ability to apply mathematical formulas and numerical reasoning to solve physics problems. For many, it involved identifying known and unknown values, selecting appropriate equations, and performing accurate calculations to arrive at a correct answer. One student explained, *"Quantitative problem-solving skills in physics mean the ability to take what we conceptually know and apply it to numbers and equations that can help us reach an exact answer."* Several emphasized that success in this area depends not only on mathematical skills but also on understanding how equations relate to physical concepts. Some students highlighted the importance of unit analysis, equation manipulation, and deriving something or reasoning through problems even when the process was not entirely familiar while solving physics problems. Others described strong quantitative skills as being able to solve homework or test problems confidently and efficiently, even without full conceptual clarity.

To support peers struggling with quantitative skills, students at all levels overwhelmingly recommended consistent practice with the course materials in physics problem-solving. The most frequent advice included doing a variety of practice problems, reviewing mistakes, and making detailed equations or formula sheets. Many students also advised attending instructor and teaching assistant's office hours, watching instructional videos (including YouTube channels), and studying with peers to gain different perspectives. Several stressed the importance of understanding units and variables, breaking problems into smaller steps, and staying organized with work. Overall, students viewed quantitative problem-solving as a trainable skill built through repetition, reflection, and focused use of resources.

From the responses given by students, it is quite clear that a majority of them possess a strong understanding of conceptual and quantitative problem-solving in the physics problem-solving context. Their definitions of conceptual and quantitative problem-solving, as well as the strategies they described using to improve on these, are often consistent with how many faculty typically described them (including in our interviews) in instructional and professional contexts.

**RQ2. How do introductory women and men students' self-assessed conceptual and quantitative problem-solving skills differ by gender across various levels of physics courses?**

We examine how women and men self-assess their conceptual and quantitative problem-solving skills in introductory algebra-based and calculus-based physics courses, using four categories: poor, fair, good, and excellent. The graphs below show the percentage of respondents in each gender group within these categories for Physics 1 courses. The graphs for Physics 2 courses are included in Appendix B. We present only the post-survey data here and the corresponding pre-survey data, are included in the Appendix C. Women are represented by orange bars and men by blue bars.

As shown in Figure 1, in algebra-based Physics 1, 43% of women and 58% of men rated their conceptual skills as good or excellent, while 60% of women and 69% of men rated their quantitative skills, similarly, showing women generally rate themselves lower. Both groups were more confident in quantitative than conceptual skills, with men increasing their quantitative self-ratings by course end compared to pre-survey (Figure A5 in Appendix C).

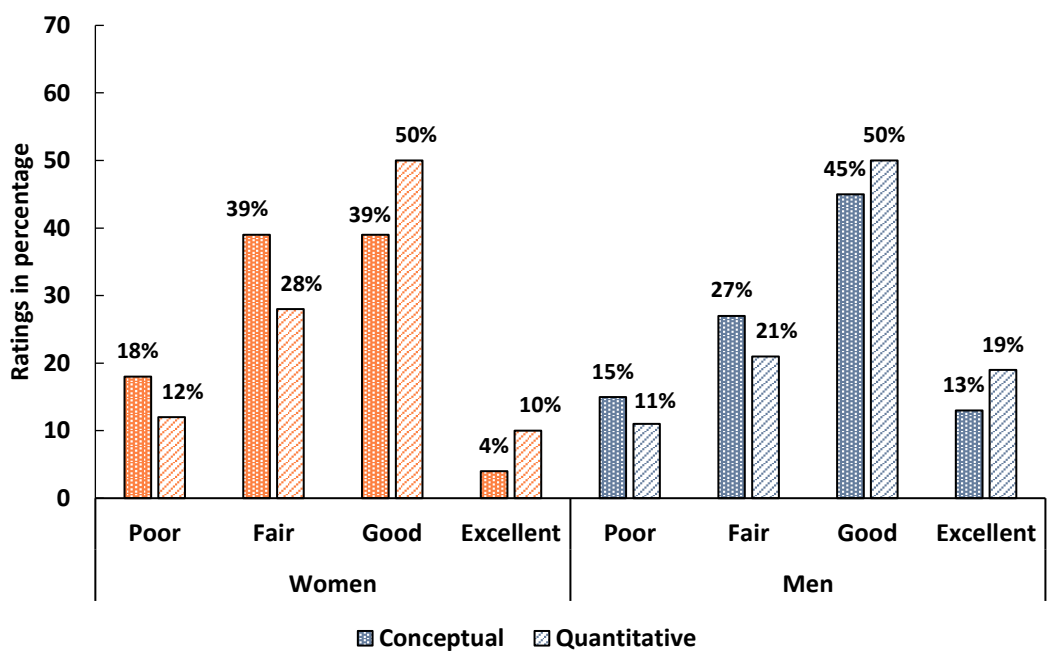

**Figure 1.** Self-assessed conceptual and quantitative problem-solving skills of women (orange bars) and men (blue bars) in algebra-based physics 1 (post survey).

A similar pattern occurred in Physics 2 with 37% of women and 57% of men rating their conceptual skills as good or excellent, while quantitative ratings were 61% for women and 66% for men (Figure A1 in Appendix B), indicating a smaller gender gap in quantitative skills. Women more often chose "Good" and men "Excellent." Both genders remained more confident in quantitative than conceptual skills, though conceptual self-ratings dropped 15–17% by the end of the course compared to pre-survey (Figure A6 in Appendix C), highlighting decreased confidence in this area.

In calculus-based Physics 1, 49% of women and 68% of men rated their conceptual skills as good or excellent, while 61% of women and 74% of men did so for quantitative skills (Figure 2). Men showed particularly high confidence in quantitative abilities, both relative to women and to men in algebra-based courses. Women's quantitative self-ratings declined from 68% to 61% by course end, indicating a drop in confidence compared to the pre-survey (Figure A7 in Appendix C).

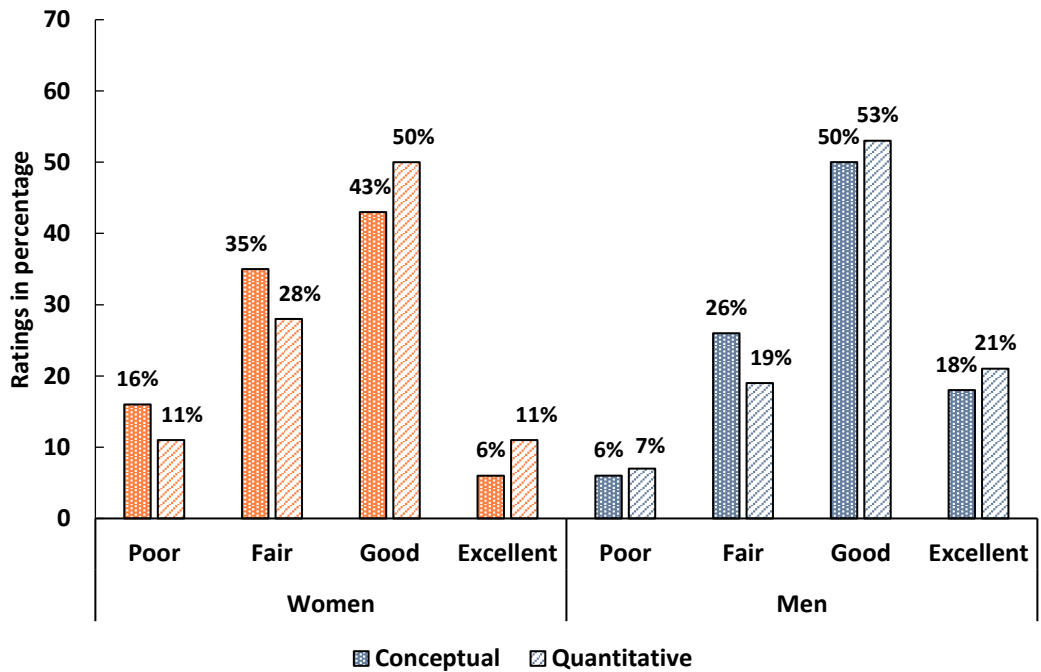

**Figure 2.** Self-assessed conceptual and quantitative problem-solving skills of women (orange bars) and men (blue bars) in calculus-based physics 1 (post survey).

In calculus-based Physics 2, 51% of women and 70% of men rated their conceptual skills as good or excellent, while 74% of women and 81% of men did so for quantitative skills (Figure A2 in Appendix B). Women showed a clear gap between conceptual and quantitative self-ratings. Both genders reported higher confidence in quantitative skills than in other introductory courses, with nearly a 10% increase from the pre-survey (Figure A8 in Appendix C). Women's conceptual ratings rose slightly, while men showed no notable change.

Overall trends across algebra and calculus-based courses show men consistently rated both skill types higher than women, with slightly higher conceptual ratings in calculus-based courses. Effect sizes along with 95% CI (shown as min and max value in Table 6) indicate women rated their skills lower than men, with gender differences in conceptual skills statistically significant in both pre- and post-survey for all introductory courses. For quantitative skills, the gender gap was statistically significant in pre-survey for all courses. Similar results were observed for post-survey in all courses except algebra-based Physics 2. The emphasis on algorithmic problem solving over conceptual reasoning in graded work may contribute to lower self-efficacy in conceptual skills, especially in women.

**Table 6.** Effect sizes given by Cohen's $d$ for the gender differences in students' self-ratings of conceptual and quantitative skills for pre and post surveys in introductory physics courses (* represents $p < 0.05$, ** represents $p < 0.01$, and *** represents $p < 0.001$, shown in bold) [107].

| | | **Introductory Courses** | **Effect Size** | **Min Value** | **Max Value** |
|---|---|---|---|---|---|
| Pre-survey | Conceptual Skills | Algebra based Physics 1 | **−0.45 *** | −0.59 | −0.31 |
| | | Algebra based Physics 2 | **−0.55 *** | −0.73 | −0.37 |
| | | Calculus based Physics 1 | **−0.27 *** | −0.43 | −0.11 |
| | | Calculus based Physics 2 | **−0.73 *** | −0.93 | −0.53 |
|  | Quantitative Skills | Algebra based Physics 1 | −0.17 * | −0.31 | −0.03 |
|  |  | Algebra based Physics 2 | **−0.42 *** | −0.60 | −0.24 |
|  |  | Calculus based Physics 1 | −0.17 * | −0.33 | −0.01 |
|  |  | Calculus based Physics 2 | −0.33 ** | −0.53 | −0.13 |
| Post survey | Conceptual Skills | Algebra based Physics 1 | **−0.42 *** | −0.58 | −0.26 |
|  |  | Algebra based Physics 2 | **−0.50 *** | −0.70 | −0.30 |
|  |  | Calculus based Physics 1 | **−0.58*** | −0.78 | −0.38 |
|  |  | Calculus based Physics 2 | **−0.50 *** | −0.74 | −0.26 |
|  | Quantitative Skills | Algebra based Physics 1 | **−0.23 ** | −0.39 | −0.07 |
|  |  | Algebra based Physics 2 | −0.15 | −0.35 | −0.05 |
|  |  | Calculus based Physics 1 | **−0.38 *** | −0.58 | −0.18 |
|  |  | Calculus based Physics 2 | **−0.33 ** | −0.55 | −0.11 |

**RQ3. To what extent do upper-level undergraduates and graduate students' self-assessments of conceptual and quantitative problem-solving skills change from the beginning to the end of the course, and how does this vary across course levels?**

We examined student self-assessment ratings in upper-level undergraduate and graduate courses. The pre-survey is represented by orange bars and post survey by blue bars. Due to the small sample size, we did not separate responses by gender for these levels. In upper-level courses, 73% of students rated their conceptual skills as good or excellent in the pre-survey comparable to 71% in the post-survey (see Figure 3). For quantitative skills, we see a similar trend as 71% of the students in the pre-survey and 70% in the post-survey rated their skills as good or excellent as observed in Figure 3. Thus, no significant differences were observed in students' ratings from the beginning to the end of the upper-level undergraduate courses. Notably, unlike in the introductory physics courses, we also did not find significant differences between the ratings of conceptual and quantitative skills.

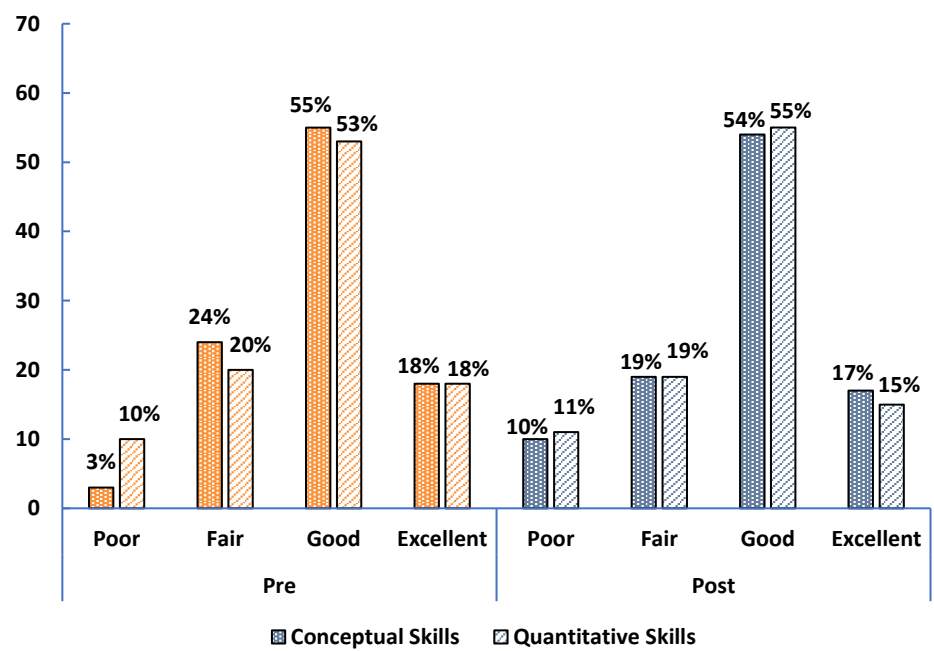

**Figure 3.** Self-assessed conceptual and quantitative problem-solving skills of upper-level undergraduate physics students (orange bars for pre-survey and blue bars for post survey).

For graduate students, 85% rated their conceptual skills as good or excellent in the pre-survey and 87% in the post-survey, while quantitative ratings rose from 75% to 83%

(Figure 4). Conceptual skills were consistently rated higher than quantitative skills. This may reflect graduate students' independent, reflective approach, integrating physics and math beyond assignments or instructor guidance, which likely supports stronger conceptual self-assessments compared to introductory students.

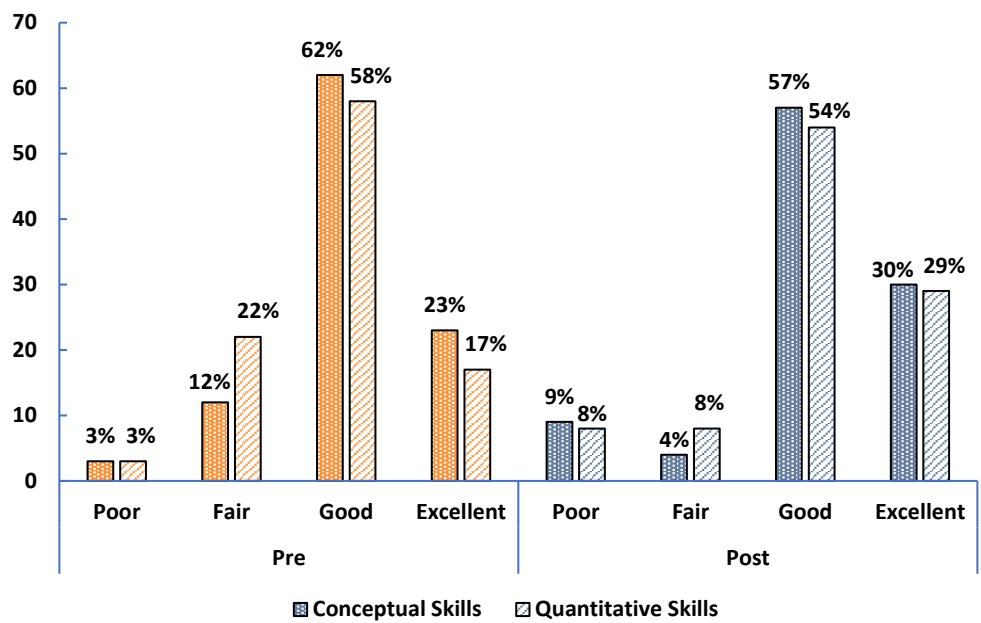

**Figure 4.** Self-assessed conceptual and quantitative problem-solving skills of graduate level physics students (orange bars for pre-survey and blue bars for post survey).

Table 7 shows effect sizes with 95% CI results comparing pre- and post-survey self-ratings for upper-level undergraduates and graduate students. The smaller graduate sample should be noted. No statistically significant changes were observed in either conceptual or quantitative skills, suggesting that advanced students may have stable self-perceptions that are less influenced by short-term instructional experiences.

**Table 7.** Effect sizes given by Cohen's *d* [107] along with 95% CI for the differences in students' self-ratings of conceptual and quantitative skills from pre to post survey in upper-level undergraduate and graduate-level physics courses. None of these effect sizes are statistically significant).

|  | Courses | Effect Size | Min Value | Max Value |
| --- | --- | --- | --- | --- |
| Conceptual Skills | Upper-Level Undergraduate | −0.08 | −0.35 | 0.19 |
|  | Graduate Level | −0.17 | −0.70 | 0.36 |
| Quantitative Skills | Upper-Level Undergraduate | 0 | −0.27 | 0.27 |
|  | Graduate Level | −0.17 | −0.68 | 0.34 |

**RQ4. How do women and men students' preferences for instructional emphasis (conceptual vs. quantitative) compare with their perceptions of what their instructors emphasize and what the course actually emphasizes across different levels of physics courses at the beginning and end of the semester?**

We surveyed students across introductory algebra-based and calculus-based physics courses, upper-level undergraduate courses, and graduate-level courses. Overall, the trends observed across these course levels were fairly consistent. Our focus was on understanding gender-based differences in how students, particularly women and men, hoped their courses would enhance their conceptual vs. quantitative problem-solving skills. In the pre-survey, students were asked what they wanted the course to improve, what they believed their instructors prioritized, and what they expected would actually happen. These questions were framed in the past tense for the post-survey to reflect their experience after completing the course.

In algebra-based physics 1, most students wanted the course to balance conceptual and quantitative emphasis or strengthen conceptual skills. Over 70% of both women and men expected their instructor to favor this balance in the pre-survey. By course end, many students, especially women (38% vs. 30% of men) felt the course focused more on quantitative skills (CQ3 in Figure 5).

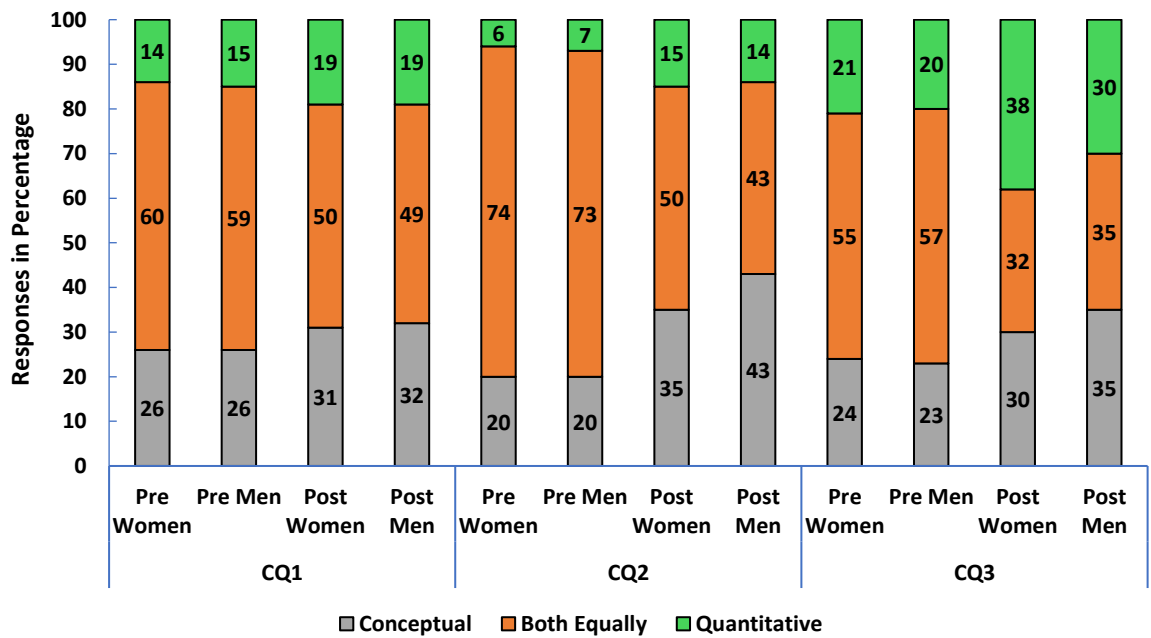

**Figure 5.** Pre- and post-course survey responses from women and men in algebra-based Physics 1: (CQ1) what students wanted the course to improve (conceptual, quantitative or both skills), (CQ2) what they thought their instructor wanted, and (CQ3) what they felt the course actually improved.

In algebra-based physics 2, nearly half of students desired a balanced emphasis, with over 65% expecting the instructor to do the same in the pre-survey. By the end, 42% of women and 29% of men reported the course improved quantitative skills more (CQ3, Figure A3 in Appendix B).

Similarly, in calculus-based physics 1, about 60% of the women and men expressed that they wanted the course to have a balance between conceptual and quantitative skills in the pre survey (CQ1 in Figure 6). Over 70% of both women and men believed the instructor also would prefer a balance between conceptual and quantitative skills in the pre-survey (CQ2 in Figure 6). By the end of the course, 39% of women and 28% of men reported that the course improved their quantitative skills more (CQ3 in Figure 6).

In calculus-based physics 2, almost half of the women and men expressed that they wanted the course to have a balance between conceptual and quantitative skills (CQ1, Figure A4 in Appendix B). About 65% of both women and men believed the instructor also would prefer a balance between conceptual and quantitative skills in the pre-survey (CQ2, Figure A4 in Appendix B). By the end of the course, 39% of women and 33% of men reported that the course improved their quantitative skills more (CQ3, Figure A4 in Appendix B). These results suggest that regardless of whether students were enrolled in algebra-based or calculus-based courses, both women and men tended to feel that the course improved quantitative skills more than conceptual understanding.

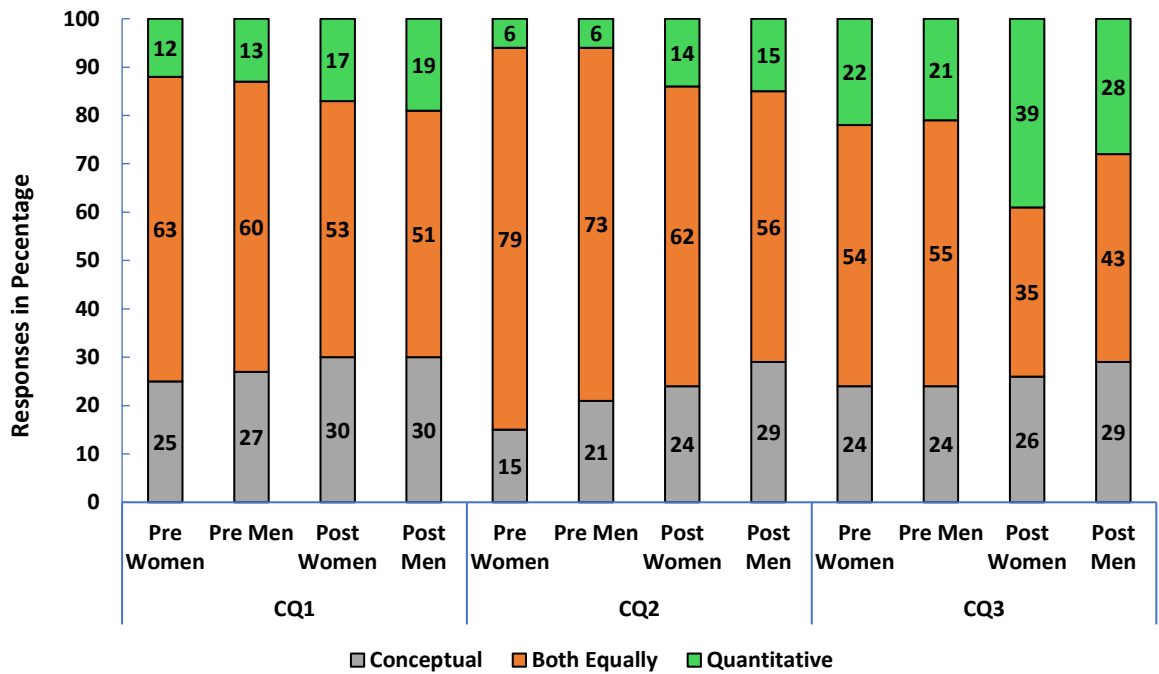

**Figure 6.** Pre- and post-course survey responses from women and men in calculus-based Physics 1: (CQ1) what students wanted the course to improve (conceptual, quantitative or both skills), (CQ2) what they thought their instructor wanted, and (CQ3) what they felt the course actually improved.

We observed similar trends among upper-level undergraduates and graduate students. In upper-level undergraduate courses, 48% of women and 20% of men felt that the course improved their quantitative skills more in the post survey (CQ3 in Figure 7), while in graduate courses, 22% of women and 15% of men reported the same (CQ3 in Figure 8).

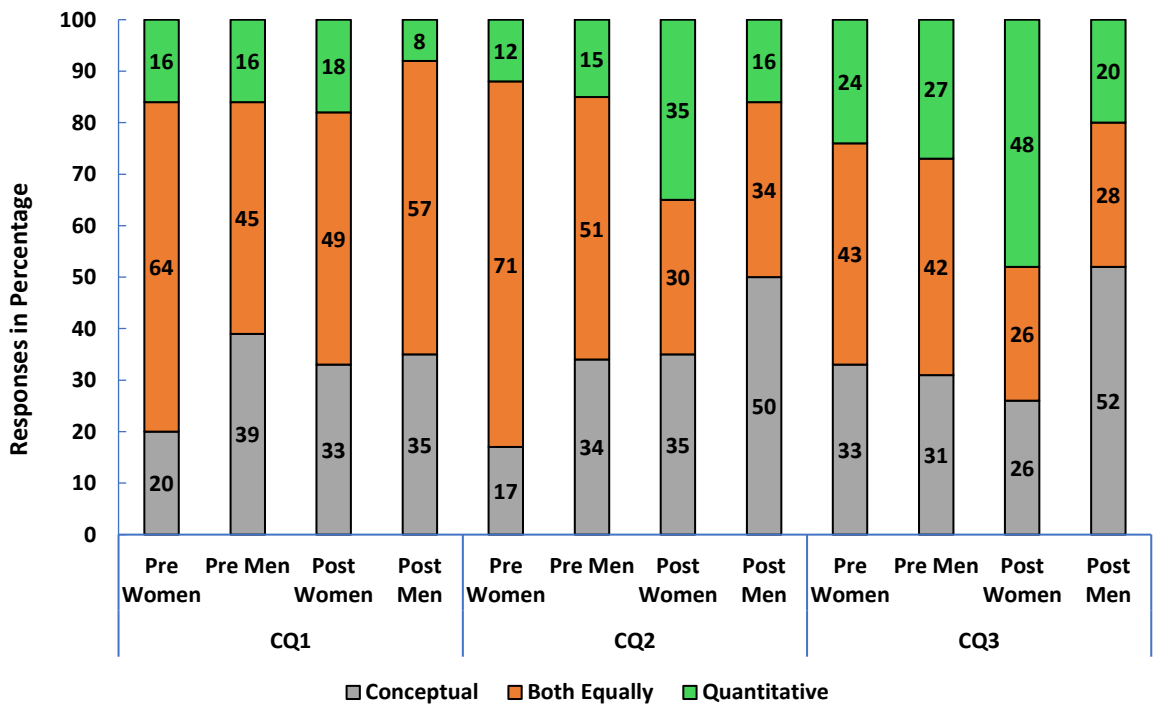

**Figure 7.** Pre- and post-course survey responses from women and men in upper-level undergraduate physics courses: (CQ1) what students wanted the course to improve (conceptual, quantitative or both skills), (CQ2) what they thought their instructor wanted, and (CQ3) what they felt the course actually improved.

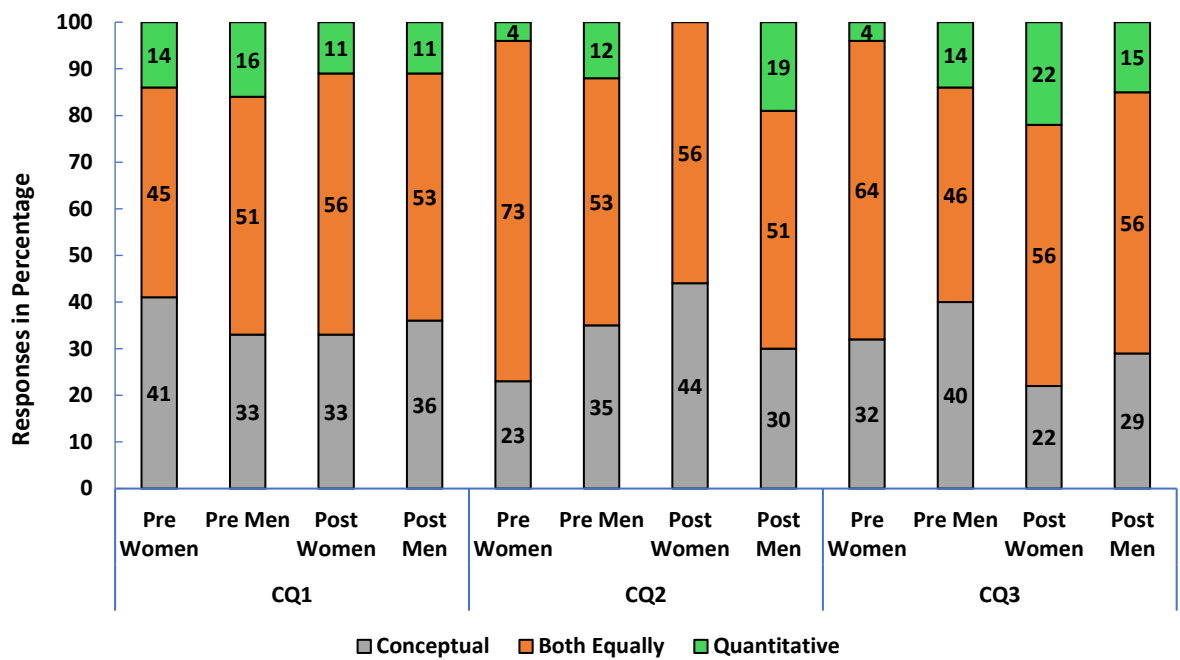

**Figure 8.** Pre- and post-course survey responses from women and men in graduate physics courses: (CQ1) what students wanted the course to improve (conceptual, quantitative, or both skills), (CQ2) what they thought their instructor wanted, and (CQ3) what they felt the course actually improved.

When asked about their instructors' priorities, only 10–15% of students in introductory courses believed the focus was primarily on quantitative skills. In contrast, 35% of women in upper-level undergraduate courses perceived a quantitative emphasis in the post-survey by the instructor (CQ2 in Figure 7). One might expect this percentage to align with what students felt actually happened in the course, but the responses suggest a disconnect. To better understand this discrepancy, we included open-ended follow-up prompts in the survey asking students to explain the reasoning behind each of their answers. This helped us capture nuanced insights into how students interpreted instructional emphasis relative to their expectations and experiences.

The reasonings reveal a noticeable gap between instructors' intentions and students' actual experiences in learning physics problem-solving skills. Many students believed that their instructors aimed to prioritize conceptual understanding, often using demonstrations, discussions, and conceptual questions. As one student noted, *"Our teacher keeps emphasizing that we need to know the why behind things."* However, students' reflections show that their quantitative skills often improved more than their conceptual skills. This was largely due to the structure of the assignments and exams, which leaned heavily on practice problems with a quantitative focus. For example, a student remarked, *"I feel like I can get problems right with math and still not understand the concepts behind it."* Others noted that while conceptual ideas were discussed, they were either rushed or confusing, leading some to default on memorizing formulas for the kinds of problems they would be graded on. One student noted, *"Demonstrations in class have helped with conceptual problem solving, but the homework and recitations are heavily based on math problems."* Another shared, *"While I did gain a better understanding of the concepts, I think this course improved my quantitative problem-solving skills more. There were a lot of practice problems, worksheets, and homework sets that focused on applying formulas, solving for unknowns, and interpreting numerical results. Over time, I became more confident in working through multi-step problems and using math as a tool to*

*support my understanding of physics. The repetition and feedback really helped me grow in that area."* These responses suggest that while instructors may have aimed to foster conceptual growth, students often gained more from the quantitatively driven aspects of the course. The difference lies in intent versus implementation: the teaching emphasized concepts, but assessment and practice prioritized equations and calculations. Thus, the students' learning outcomes were often skewed toward quantitative proficiency due to the nature of homework, recitations, and personal study habits. This highlights a pedagogical tension between instructional goals and student learning experiences.

**RQ5. How do faculty members perceive the alignment between students' instructional preferences, their own teaching goals, actual course implementation, and assessment focus in physics courses pertaining to conceptual and quantitative problem-solving?**

Faculty members who were interviewed in the study broadly agreed that conceptual and quantitative problem-solving in physics are inseparable, each enhancing the other. Conceptual reasoning allows students to interpret and frame physics problems, while quantitative aspects help carry those ideas through to a solution. As one faculty member put it, *"You need skills doing both as you get harder and harder problems…if you only do conceptual, no quantitative, you're kind of a bit restricted to certain qualitative aspects."* Another instructor emphasized that quantitative problem-solving without conceptual grounding is ineffective: *"Quantitative can't stand alone. I can't imagine teaching a purely quantitative physics class with no concepts—it wouldn't be successful."* Despite this agreement, faculty members acknowledged that the balance between the two depends heavily on the course level and student background.

At the introductory level, faculty often attempted to prioritize conceptual understanding because many students lack mathematical maturity for advanced quantitative reasoning. One instructor noted, *"You can do a lot more with conceptual at the introductory level than you can with quantitative. Students don't have the math background, so quantitative often turns into plug and chug."* Yet despite this intent, many courses still leaned heavily on quantitative assessments, which can lead students to focus on what earns points rather than what deepens understanding. As one instructor explained, *"Students are driven by what is emphasized in exams…if a course emphasizes quantitative aspects, students will lean toward that, and* vice versa.*"* Instructors also explained that student motivation varies by course type: *"For students in the algebra-based classes, who are often pre-med, they are likely focused on quantitative problem solving to do well on the MCATs. For the calculus-based classes, which usually consist of engineering and other STEM majors, I think they might prioritize understanding concepts more, as they see physics as a class they need to get through."* Another instructor pointed out, *"In algebra-based courses, the students don't give a damn about physics. They just want to get an A. In the engineering courses, the students don't give a damn about physics. They just want to get a C. But both of them [algebra-based and calculus-based introductory physics students] are more into plug and chug than the concepts. They want whatever is going to be the easiest way for them to satisfy this requirement."* However, our student survey data from these courses do not support this assumption. In fact, we find little difference between these two groups of students in how they perceived or valued conceptual and quantitative problem-solving skills. Regardless of course type, students overwhelmingly expressed a desire for a balanced approach, indicating a disconnect between faculty expectations and student perspectives.

Table 8 presents the number of faculty members who described what a typical student wants, what they themselves aim for, and what actually happens across all three levels, alongside the corresponding percentages of women and men who responded to the same items.

Table 8. Comparison of faculty responses to the students' expectations and experiences across all levels. [Note: Only 5 instructors indicated that they had taught all 3 levels].

|  |  | Faculty | | | Students (Post Survey) | |
|---|---|---|---|---|---|---|
|  |  | **Conc** | **Balanced** | **Quant** | **Women** | **Men** |
| What a typical student wants | Introductory | 2/7 | 3/7 | 2/7 | 50–59%—Bal | 48–57%—Bal |
|  | Upper-level UG | 1/7 | 6/7 | 0/7 | 49%—Bal | 57%—Bal |
|  | Graduate | 1/5 | 3/5 | 1/5 | 56%—Bal | 53%—Bal |
| What an instructor wants | Introductory | 2/7 | 5/7 | 0/7 | 50–65%—Bal | 43–59%—Bal |
|  | Upper-level UG | 0/7 | 6/7 | 1/7 | 35%—Quant, 35%—Conc | 50%—Conc |
|  | Graduate | 1/5 | 3/5 | 1/5 | 56%—Bal | 51%—Bal |
| What actually happens | Introductory | 2/7 | 2/7 | 3/7 | 38–42%—Quant | 35–46%—Bal |
|  | Upper-level UG | 3/7 | 2/7 | 2/7 | 48%—Quant | 52%—Conc |
|  | Graduate | 1/5 | 2/5 | 2/5 | 56%—Bal | 56%—Bal |

In upper-level physics courses, especially for majors, some faculty stated that they observe a shift in student motivation and maturity, and students begin to see the need for both skill sets and are more capable of integrating them. One instructor stated, *"Physics majors are very interested in understanding the concepts behind the problem solving…they also want to achieve the problem solving because they know physics builds on itself."* Still, challenges remain. Another instructor observed that even at this level, *"Students are usually better at the math…when they have trouble, it's often because they don't know where to start conceptually."* Faculty often stressed that upper-level courses should help students solidify conceptual foundations from earlier coursework and use them to tackle increasingly sophisticated quantitative challenges.

At the graduate level, the dynamic shifts again. Some interviewed faculty claimed that the graduate students tend to prioritize quantitative mastery, sometimes under the assumption that their conceptual understanding is already solid. For example, one instructor noted, *"The students are more interested in the quantitative stuff at the graduate level because they think they already have the concepts down although it's not clear what they really understand."* However, faculty often emphasized that advanced problem-solving requires embedded conceptual reasoning. For example, another instructor remarked, *"I don't think any students come away saying they're better at quantitative problem-solving after a grad class…mostly conceptual or nothing at all."* Problems at this level demand understanding of approximations, modeling choices, and limits, all of which are deeply conceptual in nature, even if embedded within complex calculations. While some instructors stated that students shift focus from quantitative (introductory level) to conceptual (upper level) and back to quantitative (graduate), our cross-sectional student data does not support this trajectory. Instead, based upon our results discussed in the preceding sections, students across all levels consistently valued both skills and sought a balanced approach to learning and problem-solving in physics.

Some instructors also expressed skepticism about whether students understand what conceptual and quantitative problem-solving means. One instructor remarked, *"What they call conceptual or quantitative may not be the same thing that I would call conceptual or quantitative."* Yet, our survey results contradict this assumption, and students consistently provided clear and accurate descriptions of these. As noted, they typically defined conceptual problem-solving as understanding and applying physical principles without relying on formulas, and quantitative problem-solving as using equations, calculations, and numerical reasoning to reach a solution. This type of consistent response from students suggests that the faculty's doubts may reflect a misalignment between instructors' perceptions and students' actual understanding.

When discussing the focus of their assessments, instructors acknowledged that introductory physics assessments may often be seen by students as heavily quantitative, though they themselves aim for a more balanced approach. As one instructor noted, *"My students would say it was quantitative, but I would say it's a balance."* They explained that while most problems have numerical or symbolic answers, solving them still requires conceptual understanding. *"To solve the quantitative questions, the students need to know the concepts…I don't think you can strip those out and create a physics exam that's just quantitative, unless it was like purely plug and chug."* In other words, even when assessments appeared to focus primarily on quantitative problem-solving to students, some interviewed instructors claimed that deep understanding of the underlying physics concepts is essential to solve them.

Another instructor mentioned that the complexity and structure of assessment tasks evolve significantly from introductory to graduate levels. At the introductory level, *"most of the problems we give to students require basically one major problem-solving step,"* which tends to fall neatly into either conceptual multiple-choice questions or straightforward quantitative problems. However, at the graduate level, assessments demand a more integrated and nuanced approach. *"You're not asking these simple multiple-choice questions…the conceptual understanding needs to be weaved into a more complex problem."* Some interviewed instructors stated that they expected students to evaluate methods, make valid approximations, and determine which terms to include, activities that reflect high-level conceptual thinking even when embedded in quantitative work. For example, one instructor noted, *"I try to put those [conceptual] things into the problem…and not have a focus on solving integrals or differential equations. That's not the main part."*

Despite these goals, interviewed instructors often felt restricted in how they design assessments, especially in large introductory courses since different instructors must be consistent across different sections of the same course. One instructor shared, *"I don't feel as though that's necessarily representative of my preferences…there are conventions to assess our students in as uniform a manner as possible across different sections."* This pressure to align, e.g., with other colleagues' exam styles, often emphasizing quantitative problem-solving, can limit their ability to incorporate conceptual problem-solving as freely as they'd like. *"We end up being a little bit in lock…the prevailing point of view values more of quantitative problem-solving over conceptual understanding."* These institutional constraints highlight a tension between instructors' pedagogical intentions and the standardized assessment practices that dominate many large introductory physics courses.

In conclusion, most faculty members who were interviewed recognized the importance of nurturing both conceptual and quantitative proficiency to develop expertise in physics problem-solving. However, the balance between them is often distorted by assessment norms and course structure. Some instructors acknowledged that students' motivation to engage with conceptual or quantitative problem-solving in physics is heavily shaped by external factors, particularly what is emphasized in assignments and exams, and ultimately, what earns them points. As a result, it is not enough for instructors to state that conceptual understanding is important or to focus on it in lectures if this is not clearly reflected in the way assignments and assessments are designed to evaluate student learning. Our analysis further suggests that students often report greater improvement in their quantitative problem-solving skills by the end of a course, largely because these are the skills they repeatedly practice through homework and exams. This indicates that students understand the importance of conceptual problem-solving in physics, but they simply follow the cues that instructors and course structures provide. Therefore, faculty members must take greater responsibility for aligning their assessment practices with their instructional goals. The challenge for instructors is to design learning experiences and evaluations that reflect the true nature of physics as a discipline which demands integration of physics and math, and conceptual and quantitative problem-solving are two sides of the same coin. Thus, providing students with meaningful opportunities to develop expertise in physics problem-solving and demonstrate reasoning and learning that integrates physics and mathematics is critical.

## 5. Conclusions

This study provides a multi-level perspective on conceptual and quantitative problem-solving in physics, spanning introductory to graduate courses. Unlike prior work focused on single courses or only students at one level, we examine both student and instructor perspectives, capturing how understanding and priorities evolve across levels. Students consistently show well-formed views on the importance of both, while faculty face constraints that can limit alignment between belief and practice. Gender differences are prominent in introductory courses, particularly in conceptual problem-solving self-assessments. Also, upper-level undergraduates and graduate students show stable self-rating. The study also highlights how students define conceptual and quantitative problem-solving, revealing potential mismatches with faculty expectations, and suggesting ways instruction can better align with student perspectives. By combining survey data, qualitative responses, and faculty interviews, the study offers a comprehensive picture of learning priorities across course levels, with implications for curriculum design, inclusive teaching, and integration of conceptual and quantitative reasoning. Overall, it provides both empirical evidence and practical guidance for aligning instruction with student learning needs.

*5.1. Student and Faculty Perspectives on Physics Problem-Solving*

This study sheds light on the nuanced ways that conceptual and quantitative problem-solving are perceived, taught, and valued in the context of physics education from introductory to graduate levels. Through a combination of student surveys and faculty interviews, we find that while both groups agree on the importance of learning to integrate physics and math and developing conceptual and quantitative problem-solving skills, there are notable gaps in perception, practice, and instructional emphasis that need to be addressed.

Across all levels, students demonstrate a consistent understanding of conceptual and quantitative problem-solving. Conceptual problem-solving is commonly described by students as the ability to understand physical principles, reason through scenarios, and explain why a phenomenon occurs without relying solely on equations. Quantitative problem-solving, on the other hand, is described as associated with the accurate and strategic application of mathematics to solve physics problems and generate solutions. Their responses suggest that students possess a thoughtful understanding of these issues, unlike faculty concerns we observed. We observe that students consistently prefer a balanced emphasis on conceptual and quantitative problem-solving, with little variation by course level, challenging the assumption that priorities shift as students advance beyond introductory courses.

We also find that students value course assessments and instructional practices that give them opportunities to engage in both aspects of problem-solving. They often noted that future students in physics courses would benefit from strategies like drawing diagrams, visualizing scenarios, attending office hours, and explaining concepts aloud showing the value of investment in building conceptual fluency. Furthermore, students reported the importance of practice, repetition, and math review to strengthen the quantitative aspects of problem-solving. These reflections may suggest that students want to learn ways to develop their physics expertise.

Faculty responses reinforce the idea that conceptual and quantitative problem-solving and reasoning are essential and inseparable in physics. Instructors articulated that conceptual aspects support quantitative accuracy, and that meaningful problem-solving requires both. However, despite these beliefs, the reality of course assessments, particularly in introductory and large enrollment courses, often skews heavily toward quantitative tasks. Institutional norms, pressure for consistency across different sections of the same course, and limited resources constrain instructors' ability to experiment with more conceptually driven assessment formats.

This mismatch between instructors' pedagogical goals, especially in introductory physics courses and what actually happens in practice, may contribute to a gap between student experience and faculty intent. For example, although instructors report that quantitative problems implicitly have conceptual thinking built into them, students may not recognize or engage with these elements if assessments do not explicitly reward them. Moreover, when students are asked to rate the emphasis of their courses, many reported that quantitative problem-solving is foregrounded regardless of instructors' attempts at balance. These responses are likely shaped by the structure of homework, quizzes and exams, which often prioritize and reward students for calculations over explicit qualitative reasoning [108-111].

These findings can be understood through the lens of expert–novice learning and epistemological framing. Our findings show that students tend to rate their quantitative skills higher than their conceptual skills in physics problem-solving, showing they are still developing expertise. From an epistemological framing perspective, students interpret "what it means to do physics" based on instructional cues and assessment structure. If an introductory physics student works on homework and exams dominated by quantitative problems in which only the final answer is valued, they are likely to focus on getting the correct answers. In this context, they are likely to prioritize the technical role of math, performing calculations, and getting numerical solutions over understanding the underlying principles and concepts. These types of courses would provide little incentive to engage in the structural role of math in physics, which involves using mathematics to model relationships and reason about physical systems. Since students focus on what is explicitly rewarded, assessment practices can lead them to frame physics as primarily quantitative, potentially overlooking conceptual aspects even when instructors believe they have embedded them implicitly. The synergistic frameworks discussed in the introductory section help explain how assessment design can shape whether students engage with math as a tool for calculation or as a means of learning and building deep conceptual understanding of physics.

*5.2. Gender Disparities in Conceptual Skill Perceptions*

Gender differences in perceptions are evident in our findings, with women consistently rating their conceptual problem-solving skills lower than men in both algebra and calculus-based courses. We also observed that women were more likely to say that the courses actually focused on quantitative problem-solving in the post-survey compared to men across all levels. Also, 35% of women in upper-level undergraduate courses perceived that their instructors prioritized quantitative problem-solving. One possible explanation for these gendered findings is lower physics self-efficacy and identity as well as the influence of stereotype threat women experience in physics. Since women in physics often have been subjected to the stereotype that physics requires brilliance and men are more naturally suited for the discipline, their physics self-efficacy and identity can suffer even when the actual performance is comparable [112]. If quantitative problem-solving is emphasized in course assessment, women are more likely to perceive that their conceptual problem-solving skills in physics are not adequate and rate themselves even lower on this dimension as we observed in this research. In addition, classroom environments and instructional practices may unintentionally reinforce these gendered perceptions. Social dynamics in group work and lack of role models may further contribute to these self-assessments. Previous research suggests that women are more likely to drop physical science and engineering majors, even when they have higher grades than men [87]. Also, according to data from the American Institute of Physics (AIP) [113], only about 27% of AP Physics C test-takers are women. These differences suggest that underrepresentation of women in physics begins early. Due to all these issues, if conceptual problem-solving is not explicitly modeled and emphasized in physics courses, women are more likely to rate themselves lower on it.

In summary, our findings point to a shared recognition of the importance of both conceptual and quantitative aspects of physics problem-solving, but also to a disconnect between instructional intent, assessment practices, and student perceptions. Instructor interviews suggest that institutional and instructional structures may sometimes limit the realization of the instructor's goal. Bridging this gap will require deliberate effort on the part of departments and instructors including rethinking assessment formats and creating more opportunities for students to explicitly demonstrate conceptual reasoning alongside quantitative proficiency and rewarding them for it.

## 6. Recommendations and Instructional Implications

This study advances the understanding of how conceptual and quantitative problem-solving and reasoning are perceived and valued in physics education. By combining large-scale survey data with faculty interviews, it demonstrates that students across course levels, from introductory to graduate classes, generally possess a meaningful understanding of both aspects of problem-solving and consistently advocate for balanced instructional emphasis. Faculty share this view in principle but often face contextual and institutional barriers that prevent perfect alignment between belief and practice. The findings challenge faculty narratives about students' views on physics problem-solving and instead portray students as learners who understand the value of engaging with both conceptual and quantitative aspects.

The results have several important implications for teaching and curriculum design. First, instructors can benefit from these findings and engage in developing students' problem-solving skills through rewarding students for using, e.g., ASPIRE heuristic, that emphasizes systematic approach to problem-solving and assigning problems that focus on conceptual solutions. Instructors can model effective problem-solving and help students via structured discussions, reflective exercises, and other activities that build on students' prior knowledge. Second, our findings suggest that even when instructors value conceptual aspects of problem-solving, practical constraints, e.g., consistency across different sections of the same course and grading constraints, often make it difficult for them to assign problems, e.g., focusing on conceptual solutions or reward students for effective problem-solving strategies using ASPIRE heuristic in assessments. Embedding conceptual tasks within quantitative problems such as short written explanations, conceptual checkpoints, or diagrams that are only graded for completeness can make conceptual thinking more explicit without adding additional grading burden.

Third, the study highlights gender differences in students' ratings and perceptions of physics problem-solving, emphasizing the need for inclusive and equitable classrooms. Instructors should ensure that women and men alike have equal opportunities to ask questions, engage in discussions and learn to integrate conceptual and quantitative aspects of problem-solving. Awareness of stereotype threats and proactive strategies to create a safe, collaborative environment are essential for maximizing student participation and confidence [114].

Finally, our findings show that students' appreciation for both conceptual and quantitative problem-solving in physics remains consistent across levels. Thus, the curricula should intentionally reinforce the connection between the two rather than assuming it will naturally develop in higher-level courses. In particular, instructors should explicitly emphasize the connection between conceptual and quantitative problem-solving, giving explicit importance to concepts before focusing on derivations or equations in physics problems. Supporting students' metacognition is critical and can be achieved using a cognitive apprenticeship model [76], in which instructors first model thinking processes, provide scaffolding for productive practice in developing conceptual and quantitative problem-solving and reasoning skills, and gradually remove support to encourage independent learning and transfer of knowledge.

Future research should investigate how instructional interventions, such as collaborative problem-solving and reflective discussions, can foster equitable and conceptually rich learning environments throughout the physics curriculum. Ultimately, developing expertise and fostering meaningful problem-solving in physics is not a question of choosing between conceptual or quantitative focus, but about creating learning environments where the two are deeply integrated and equally valued explicitly, reflecting the true nature of how physicists solve problems.

**Author Contributions:** Conceptualization, C.S.; Methodology, A.G. and C.S.; Formal Analysis, A.G.; Data Curation, A.G.; Writing—original draft, A.G.; Writing—review and editing, A.G. and C.S.; Visualization, A.G. and C.S.; Supervision, C.S. The first author coded the faculty interview data and both authors discussed the code. Only the first author compiled all data for all figures and tallied the data for Table 8. All authors have read and agreed to the published version of the manuscript.

**Funding:** This research received no external funding.

**Institutional Review Board Statement:** The study was conducted in accordance with the Declaration of Helsinki and approved by the University of Pittsburgh Institutional Review Board (IRB PRO15030412, 15 April 2015).

**Informed Consent Statement:** Written informed consent requirement was waived by the university's Institutional Review Board. Verbal consent was obtained from each participant for interview data.

**Data Availability Statement:** The raw data presented in this study are not available as per institutional IRB policy.

**Conflicts of Interest:** The authors declare no conflicts of interest.

## Appendix A. *Faculty Interview Questions*

The following questions were posed to faculty members during semi-structured interviews, in which follow-up questions were posed based on the flow of conversation. Faculty members were also asked to explain the reasoning behind their responses to each question.

1. What level physics courses have you taught before?
    a. Introductory
    b. Upper level
    c. Graduate level
    d. All three levels

Answer the following questions (2–16) if you have taught a course at a particular level (introductory, undergraduate upper-level, or graduate level) and explain your reasoning for your specific choice:

2. How did you want an introductory level course to improve a typical student's conceptual understanding of physics vs. quantitative problem-solving?
   a. Mostly conceptual understanding
   b. More conceptual understanding than quantitative problem-solving, but I want both to improve
   c. Both Equally
   d. More quantitative problem-solving than conceptual understanding, but I want both to improve
   e. Mostly quantitative problem-solving

3. How do you think a typical student wanted an introductory level course to improve their conceptual understanding of physics vs. quantitative problem-solving?
   a. Mostly conceptual understanding
   b. More conceptual understanding than quantitative problem-solving
   c. Both Equally
   d. More quantitative problem-solving than conceptual understanding
   e. Mostly quantitative problem-solving

4. How do you think an introductory level course improved a typical students' conceptual understanding of physics vs. quantitative problem-solving skills?
   a. Mostly conceptual understanding
   b. More conceptual understanding than quantitative problem-solving
   c. Both Equally
   d. More quantitative problem-solving than conceptual understanding
   e. Mostly quantitative problem-solving

5. How did you want an undergraduate upper-level course to improve your students' conceptual understanding of physics vs. quantitative problem-solving?
   a. Mostly conceptual understanding
   b. More conceptual understanding than quantitative problem-solving, but I want both to improve
   c. Both Equally
   d. More quantitative problem-solving than conceptual understanding, but I want both to improve
   e. Mostly quantitative problem-solving

6. How do you think a typical student wanted an undergraduate upper-level course to improve their conceptual understanding of physics vs. quantitative problem-solving?
   a. Mostly conceptual understanding
   b. More conceptual understanding than quantitative problem-solving
   c. Both Equally
   d. More quantitative problem-solving than conceptual understanding
   e. Mostly quantitative problem-solving

7. How do you think an undergraduate upper-level course improved a typical students' conceptual understanding of physics vs. quantitative problem-solving skills?
   a. Mostly conceptual understanding
   b. More conceptual understanding than quantitative problem-solving
   c. Both Equally
   d. More quantitative problem-solving than conceptual understanding
   e. Mostly quantitative problem-solving

8. How did you want a graduate level course to improve your students' conceptual understanding of physics vs. quantitative problem-solving?

a. Mostly conceptual understanding
   b. More conceptual understanding than quantitative problem-solving, but I want both to improve
   c. Both Equally
   d. More quantitative problem-solving than conceptual understanding, but I want both to improve
   e. Mostly quantitative problem-solving
9. How do you think a typical student wanted a graduate level course to improve their conceptual understanding of physics vs. quantitative problem-solving?
   a. Mostly conceptual understanding
   b. More conceptual understanding than quantitative problem-solving
   c. Both Equally
   d. More quantitative problem-solving than conceptual understanding
   e. Mostly quantitative problem-solving
10. How do you think a graduate level course improved a typical students' conceptual understanding of physics vs. quantitative problem-solving skills?
    a. Mostly conceptual understanding
    b. More conceptual understanding than quantitative problem-solving
    c. Both Equally
    d. More quantitative problem-solving than conceptual understanding
    e. Mostly quantitative problem-solving
11. A typical student wants more conceptual focus than quantitative focus in an introductory level physics course:
    a. Strongly Disagree
    b. Disagree
    c. Slightly Disagree
    d. Neutral
    e. Slightly Agree
    f. Agree
    g. Strongly Agree
12. A typical student wants conceptual and quantitative aspects of physics to be balanced in an introductory level physics course:
    a. Strongly Disagree
    b. Disagree
    c. Slightly Disagree
    d. Neutral
    e. Slightly Agree
    f. Agree
    g. Strongly Agree
13. A typical student wants more conceptual focus than quantitative focus in an undergraduate upper-level physics course:
    a. Strongly Disagree
    b. Disagree
    c. Slightly Disagree
    d. Neutral
    e. Slightly Agree
    f. Agree
    g. Strongly Agree

14. A typical student wants conceptual and quantitative aspects of physics to be balanced in an undergraduate upper-level physics course:
    a. Strongly Disagree
    b. Disagree
    c. Slightly Disagree
    d. Neutral
    e. Slightly Agree
    f. Agree
    g. Strongly Agree
15. A typical student wants more conceptual focus than quantitative focus in a graduate level physics course:
    a. Strongly Disagree
    b. Disagree
    c. Slightly Disagree
    d. Neutral
    e. Slightly Agree
    f. Agree
    g. Strongly Agree
16. A typical student wants conceptual and quantitative aspects of physics to be balanced in a graduate level physics course:
    a. Strongly Disagree
    b. Disagree
    c. Slightly Disagree
    d. Neutral
    e. Slightly Agree
    f. Agree
    g. Strongly Agree
17. What fraction of a physics class do you think would value conceptual over quantitative focus? [at each level faculty member had taught]

    a. 25% b. 50% c. 75% d. 100%
18. What were your assessments focused on? [at each level faculty member had taught]
    a. More conceptual understanding of physics
    b. More quantitative problem solving
    c. Balance of conceptual understanding and quantitative problem solving

**Appendix B.** *Graphs for Algebra-Based and Calculus-Based Physics 2 Courses*

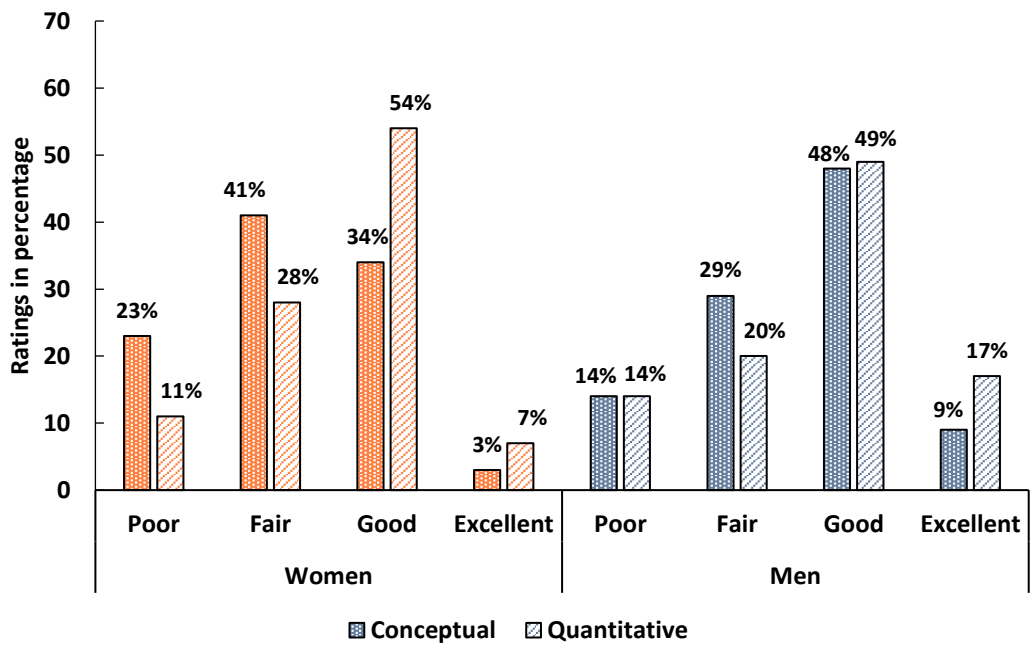

**Figure A1.** Self-assessed conceptual and quantitative problem-solving skills of women (orange bars) and men (blue bars) in algebra-based physics 2 (post survey).

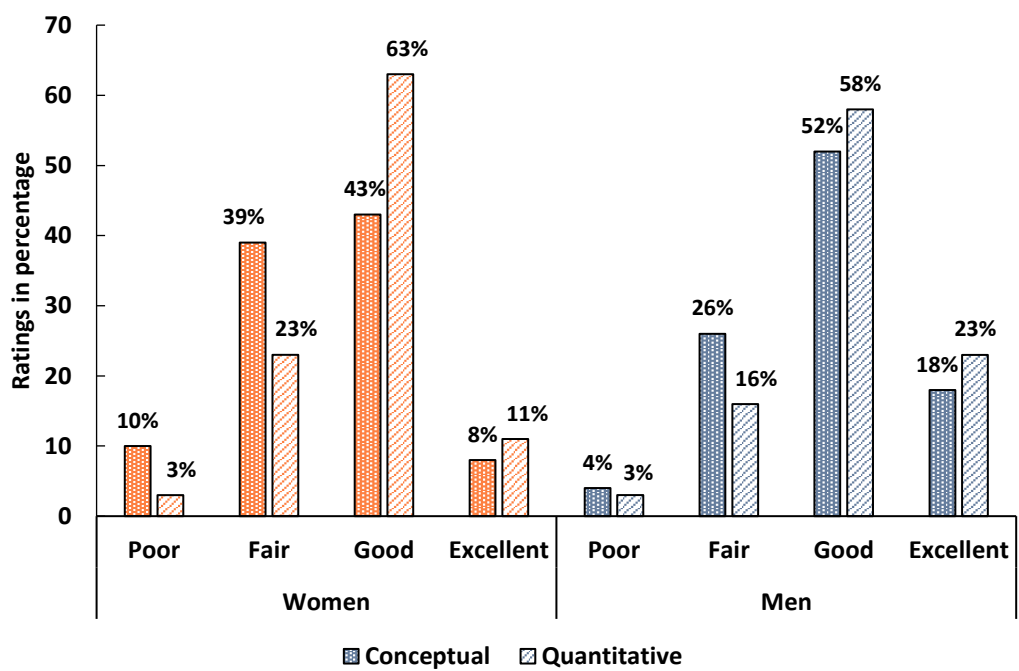

**Figure A2.** Self-assessed conceptual and quantitative problem-solving skills of women (orange bars) and men (blue bars) in calculus-based physics 2 (post survey).

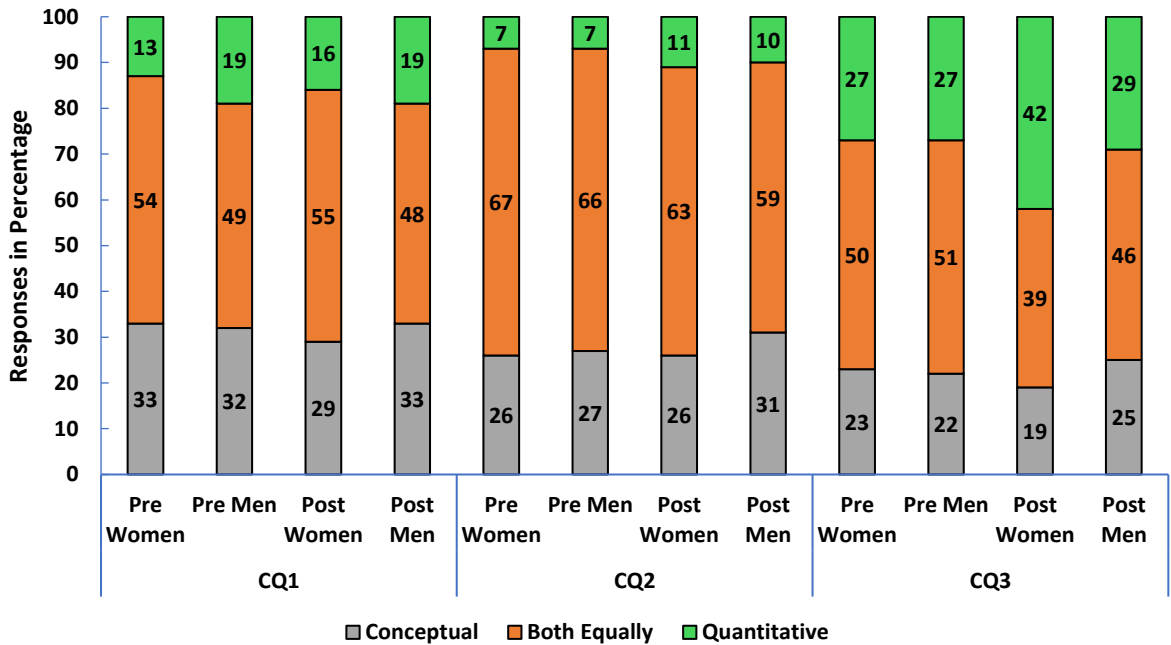

**Figure A3.** Pre- and post-course survey responses from women and men in algebra-based Physics 2: (CQ1) what students wanted the course to improve (conceptual, quantitative, or both skills), (CQ2) what they thought their instructor wanted, and (CQ3) what they felt the course actually improved.

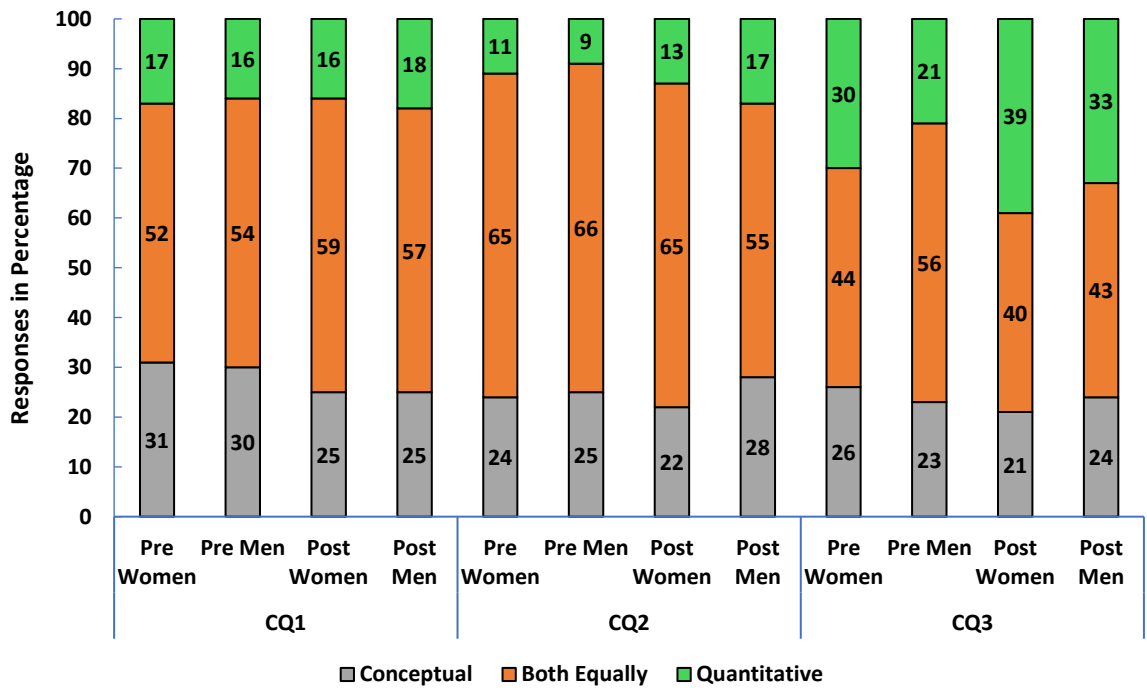

**Figure A4.** Pre- and post-course survey responses from women and men in calculus-based Physics 2: (CQ1) what students wanted the course to improve (conceptual, quantitative, or both skills), (CQ2) what they thought their instructor wanted, and (CQ3) what they felt the course actually improved.

**Appendix C.** *Graphs for Self-Assessment of Conceptual and Quantitative Problem-Solving Skills in Introductory Physics Courses (Pre-Survey)*

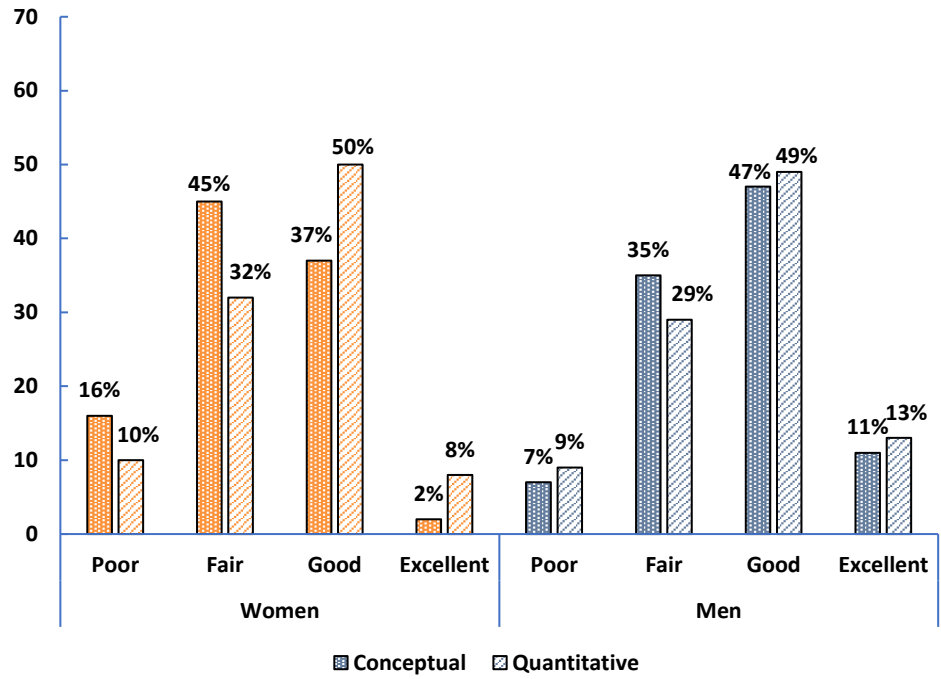

**Figure A5.** Self-assessed conceptual and quantitative problem-solving skills of women (orange bars) and men (blue bars) in algebra-based physics 1.

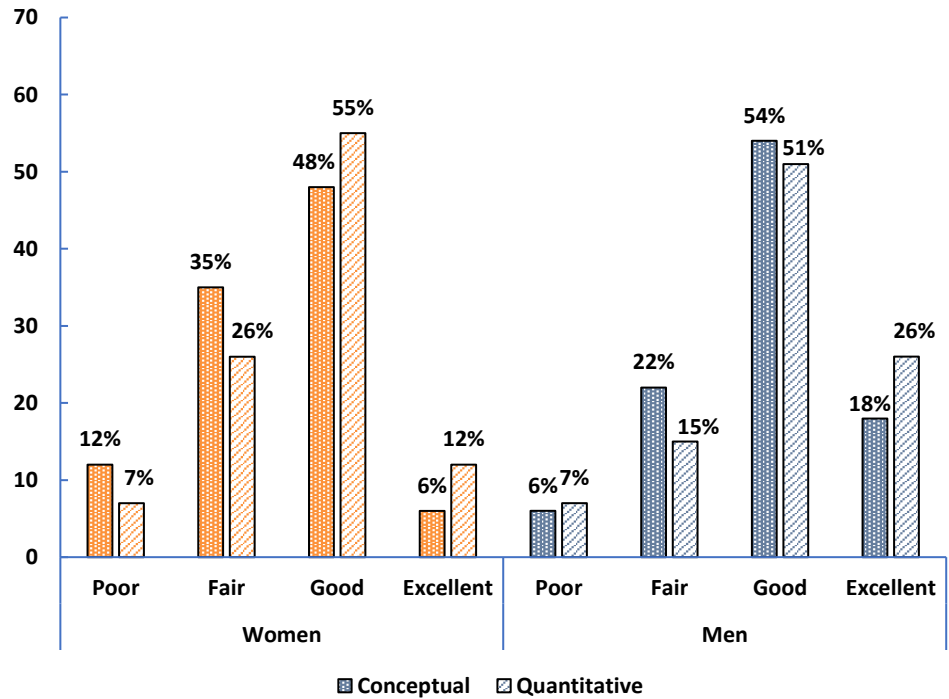

**Figure A6.** Self-assessed conceptual and quantitative problem-solving skills of women (orange bars) and men (blue bars) in algebra-based physics 2.

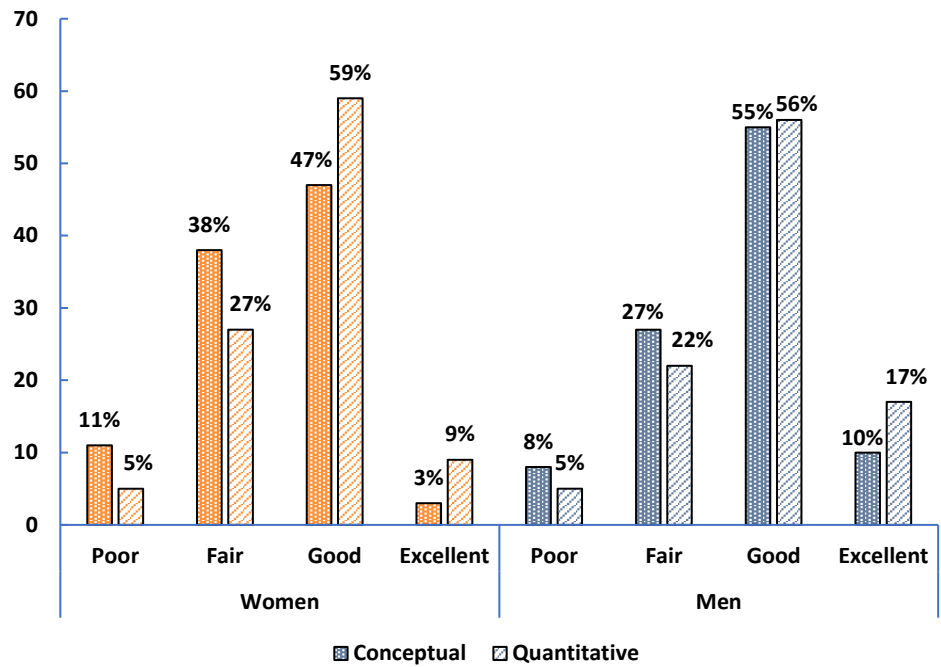

**Figure A7.** Self-assessed conceptual and quantitative problem-solving skills of women (orange bars) and men (blue bars) in calculus-based physics 1.

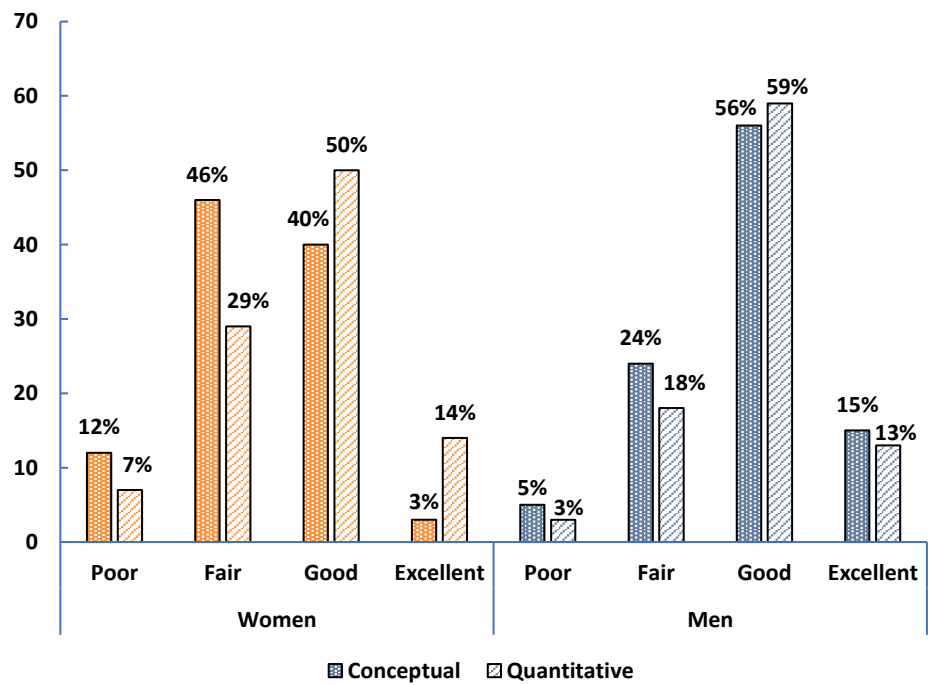

**Figure A8.** Self-assessed conceptual and quantitative problem-solving skills of women (orange bars) and men (blue bars) in calculus-based physics 2.